\documentclass[twocolumn,showpacs,preprintnumbers,amsmath,amssymb]{revtex4}
\usepackage[dvips]{graphicx} 
\begin{document}

\preprint{IMAFF-RCA-03-08}
\title{Dark Energy Accretion onto black holes in a cosmic scenario}
\author{Prado Mart\'{\i}n-Moruno$^{1}$, Az-Eddine L. Marrakchi$^{2}$,
Salvador Robles-P\'{e}rez$^{1}$ and Pedro F. Gonz\'alez-D\'{\i}az$^{1}$}
\affiliation{$^{1}$Colina de los Chopos, Centro de F\'{\i}sica ``Miguel Catal\'{a}n'',
Instituto de Matem\'{a}ticas y F\'{\i}sica Fundamental,\\
Consejo Superior de Investigaciones Cient\'{\i}ficas, Serrano 121, 28006
Madrid (SPAIN). \\
$^{2}$Laboratoire de Physique Th\'{e}orique et Appliqu\'{e}e, Facult\'{e}
des Sciences - Dhar El Mahraz, Universit\'{e} Sidi Mohamed Ben Abdellah, \\
B.P. 1796 F\'{e}s - Atlas, F\'{e}s (MOROCCO).}
\date{\today}

\begin{abstract}
In this paper we study the accretion of dark energy onto a black hole in the
cases that dark energy is equipped with a positive cosmological constant and
when the space-time is described by a Schwarzschild-de Sitter metric. It is
shown that, if confronted with current observational data, the results
derived when no cosmological constant is present are once again obtained in
both cases.
\end{abstract}

\maketitle

\section{Introduction}

In the last few years a large number of models have been proposed to
describe the current accelerated expansion of our Universe \cite%
{Copeland:2006wr}. The consideration of these new models entails the
appearance of new and surprising phenomena, such as those named big rip \cite%
{Caldwell:2003vq}, big trip \cite{GonzalezDiaz:2004vv} or big hole \cite%
{MartinMoruno:2006mi,Yurov:2006we,MartinMoruno:2007se}. One of the most
popular models describing the accelerated expansion state is the
quintessence model \cite{GonzalezDiaz:2004eu}. In this scenario one
considers a time dependent scalar field which can be interpreted as an
homogeneous and isotropic fluid with an equation of state, $p=w\rho$ (where
units, $c=G=1$, are used), in which $p$ and $\rho$ are the pressure and
energy density, respectively, and $w$ is a constant which current
observations make to look close to $-1$, but with a given bias toward the
phantom regime \cite{Caldwell:1999ew}, $w<-1$. The latter case corresponds
to a "super-accelerated" expansion of the Universe, until it finally reaches
a big rip singularity \cite{Caldwell:2003vq}. If present wormholes accrete
that phantom energy, then their sizes increase so rapidly that can
eventually engulf the universe itself, so allowing for cosmic time traveling
\cite{GonzalezDiaz:2004vv}. On the other hand, present black holes accreting
dark energy with $w>-1$ could in principle eventually engulf the universe in
a process that has been dubbed big hole [6]. In this case, nevertheless,
current observational data prevent such a process from occurring \cite%
{MartinMoruno:2006mi}.

However, a question which still remains unanswered is what happens in the
case in which the cosmic expansion is caused by a certain dark stuff when a
positive cosmological constant is involved at. In that case, could the
accretion of dark energy onto black holes be so fast that the Universe would
eventually undergo a big hole phenomenon?. In order to answer that question,
two different ways can be followed. The first one is developed in sec. II,
where a cosmological constant is introduced in the Friedmann equations,
taking for the mass rate of the black hole the expression obtained from a
non-static Schwarzschild metric \cite{MartinMoruno:2006mi}. In that case, if
we consider an homogeneous and isotropic model with a black hole described
by a Schwarzschild metric, no big hole phenomenon would happen \cite%
{MartinMoruno:2007se}. In the second way Friedmann equations with no
cosmological constant are considered, using the Schwarzschild-de Sitter
metric to describe the space-time. This allows us to study in sec. III the
accretion procedure in a cosmic scenario, where the universe is no longer
asymptotically flat but asymptotically de Sitter. Finally, in sec. IV, some
conclusions are drawn and further comments added.

\section{Accretion of dark energy with $\Lambda$.}

Let us consider first a dark energy model described by a perfect fluid with
an equation of state given by $p=w\rho$, where $w$ is a constant parameter.
Integrating the expression for the cosmic energy conservation, $\mathrm{d}%
\rho=-3(p+\rho)\mathrm{d}a/a$, where $a(t)$ is the scale factor, we obtain
\begin{equation}  \label{uno}
\rho=\rho_0(a(t)/a_0)^{-3(1+w)}.
\end{equation}
In that case, the Friedmann equation for a flat universe with a cosmological
constant reads,
\begin{equation}  \label{dos}
\left(\frac{\dot{a}}{a}\right)^2=\lambda+Ca^{-3(1+w)},
\end{equation}
in which the overhead dot denotes the derivative with respect to time, $%
\lambda=\Lambda/3>0$ and $C=8\pi\rho_0/(3a_0^{-3(1+w)})$, with $a_0\equiv
a(t_0)$, being $t_0$ the time at the current epoch. Eq. (\ref{dos}) can be
integrated to give the time evolution of the scale factor \cite%
{GonzalezDiaz:2004eu,BouhmadiLopez:2004me}. We get
\begin{widetext}
\begin{equation}\label{tres}
a(t)=\left(\frac{C}{4D\lambda}\right)^{1/[3(1+w)]}\left(e^{\frac{3}{2}(1+w)\lambda^{1/2}(t-t_0)}-De^{-\frac{3}{2}(1+w)\lambda^{1/2}(t-t_0)}\right)^{2/[3(1+w)]},
\end{equation}
\end{widetext}
where
\begin{equation}  \label{cuatro}
D=\frac{(\lambda+Ca_0^{-3(1+w)})^{1/2}-\lambda^{1/2}}{%
(\lambda+Ca_0^{-3(1+w)})^{1/2}+\lambda^{1/2}},
\end{equation}
from which it can be easily checked that, $0<D<1$.

As is well known, the general expression for the black hole mass rate in the
case where the black hole accretes dark energy in a non-static Schwarzschild
space, is given for an asymptotic observer by \cite{MartinMoruno:2006mi},
\begin{equation}  \label{cinco}
\dot{M}=4\pi AM^2(p+\rho),
\end{equation}
where $A$ is a constant. This rate equation can be integrated in any
homogeneous and isotropic model, giving \cite{MartinMoruno:2007se}
\begin{equation}  \label{M4DH}
M(t)=\frac{M_0}{1+AM_0\left[H(t)-H_0\right]}
\end{equation}
which, taking into account Eq. (\ref{tres}), can be re-expressed as
\begin{equation}  \label{seis}
M(t)=M_0\left[1-\frac{2\lambda^{1/2}DM_0A}{1-D}\frac{e^{3(1+w)%
\lambda^{1/2}(t-t_0)}-1}{e^{3(1+w)\lambda^{1/2}(t-t_0)}-D}\right]^{-1}.
\end{equation}
If the Universe is filled with a phantom fluid, for which $w<-1$, the scale
factor given by Eq. (\ref{tres}) diverges at the so-called big rip
singularity \cite{Caldwell:2003vq}, which takes place at a finite time $%
t_{br}$ in the future given by
\begin{equation}  \label{siete}
t_{br}=t_0+\frac{|\mathrm{Ln}D|}{3(|w|-1)\lambda^{1/2}}.
\end{equation}
From Eq. (\ref{seis}), it can be seen that the black hole mass can be
written as, $M=M_0[1-F(t)]^{-1}$, where $F(t)$ is a decreasing function of
time and $F(t_0)=0$, i. e., $F(t)$ is a negative function. Then, the black
hole mass will decrease progressively until it vanishes at the Big Rip, for
any $M_0$. Therefore, any black hole in a phantom dominated universe tends
to disappear at the big rip singularity, a result that was already derived
by Babichev \textit{et al.} [10] for the static case with $\Lambda =0$.

On the other hand, if a fluid with $w>-1$ is considered, then the scale
factor of the Universe would steadily increase forever (see Eqs. (\ref{tres}%
) and (\ref{cuatro})). In that case, the function $F(t)$ becomes an
increasing function of time with a finite limit given by, $%
\gamma=2\lambda^{1/2}M_0DA/(1-D)$, when $t \rightarrow \infty$. Then, the
Universe will suffer a big hole phenomenon if $\gamma>1$. Expliciting the
expression for $D$ leads to the condition
\begin{equation}
M_0A\left[(\lambda+8\pi\rho_0/3)^{1/2}-\lambda^{1/2}\right]>1.
\end{equation}
However, from the WMAP data, it follows that $H_0\sim 10^{-26}(\mathrm{meters%
})^{-1}$, and making use of Eq. (\ref{dos}), it can be estimated that $%
(\lambda+8\pi\rho_0/3)^{1/2}\sim 10^{-26}$, and generally, $\gamma\sim
M_0\left(10^{-26}-\lambda^{1/2}\right)<M_010^{-26} \ll 1$, as it is shown in
\cite{MartinMoruno:2006mi,MartinMoruno:2007se}. Therefore, there could exist
no black holes in the Universe with a initial mass so big as to produce a
big hole phenomenon. The conclusion can be moreover drawn that for a black
hole to produce a big hole phenomenon in a universe with a positive
cosmological constant it is necessary that originally it be even bigger than
those that could produce a big hole in a universe without cosmological
constant.

\section{Accretion onto Schwarzschild-De Sitter.}

In the Babichev \textit{et al.} model \cite{Babichev:2004yx}, and its
generalization for the non-static metric \cite{MartinMoruno:2006mi}, the
black hole was described by means of the Schwarzschild metric (its
non-static generalization), immersed in dark energy depicted as a perfect
fluid. In the previous section we have inserted a positive cosmological
constant in the Friedmann equations. We had therefore a universe filled with
an "effective fluid" containing both dark energy and a cosmological
constant, where a static or non-static Schwarzschild black hole was immersed.

In the present section it will be instead assumed that the stuff which
produces the expansion of the Universe is a dark energy fluid alone, and the
contribution of the cosmological constant is taken into account through the
space-time metric itself; that is, we consider now a Schwarzschild-de Sitter
metric to describe that space-time,
\begin{equation}  \label{ocho}
\mathrm{d}s^2=-\frac{\Delta}{r^2}\mathrm{d}t^2+r^2\left(\frac{\mathrm{d}r^2}{%
\Delta}+\mathrm{d}\theta^2+\sin^2\theta\mathrm{d}\phi^2\right),
\end{equation}
where,
\begin{equation}  \label{nueve}
\Delta=r^2\left(1-\frac{r^2\Lambda}{3}-\frac{2M}{r}\right).
\end{equation}
The possible horizons of this space would be the zeros of the $\Delta$
function. One can see that if $3M\Lambda^{1/2}>1$ there is only a real
solution which is negative; if $3M\Lambda^{1/2}=1$, there is one negative
real solution and two equal positive real solutions; and if $%
3M\Lambda^{1/2}<1$, then there is one negative real solution and two
different positive real solutions. In the latter case, one can provide these
two positive real solutions with the physical meaning of two horizons \cite%
{Gibbons:1977mu}. So,in that case, there is a black hole horizon $r_{bh}$
and a cosmological horizon $r_{c}$, defined by
\begin{equation}  \label{diez}
r_{bh}=\frac{2}{\Lambda^{1/2}}\cos\left[\frac{\arccos(-3M\Lambda^{1/2})}{3}+%
\frac{4\pi}{3}\right]
\end{equation}
and
\begin{equation}  \label{once}
r_{c}=\frac{2}{\Lambda^{1/2}}\cos\left[\frac{\arccos(-3M\Lambda^{1/2})}{3}%
\right].
\end{equation}

If the black hole mass $M$ increases, the size of the black hole horizon
would also increase whereas that of the cosmological horizon would decrease,
until a limiting case in which $M$ takes such a value that $3M\Lambda^{1/2}=1
$, where the two horizons reduce to just one horizon, so disappearing the
physical space between them.

Now, let us assume that the space is filled by dark energy which is
described by a perfect fluid with an energy-momentum tensor given by,
\begin{equation}  \label{doce}
T_{\mu\nu}=(p+\rho)u_{\mu}u_{\nu}-pg_{\mu\nu},
\end{equation}
where $p$ and $\rho$ are the pressure and energy density, respectively, and $%
u_{\mu}$ is the four-velocity, such that $u_{\mu}u^{\mu}=-1$.

We consider, such as it was made in Refs. \cite{Babichev:2004yx} and \cite%
{MartinMoruno:2006mi}, the time component of the conservation law of the
energy-momentum tensor, $T^{\nu}_{0;\nu}=0$, and derive the following
equation
\begin{equation}  \label{trece}
\partial_rT^r_0+\frac{2}{r}T^r_0+f(r,t)=0,
\end{equation}
where
\begin{equation}  \label{catorce}
f(r,t)=\partial_0T^0_0+\frac{\dot\Delta}{2\Delta}\left(T^r_r-T^0_0\right).
\end{equation}
On the other hand, the projection onto the four-velocity of the
energy-momentum tensor conservation law gives,
\begin{equation}  \label{quince}
\frac{\partial_r\rho}{p+\rho}+\frac{\partial_ru}{u}+\frac{2}{r}+g(r,t)=0,
\end{equation}
with $u=u^r$ and
\begin{equation}  \label{dieciseis}
g(r,t)=\frac{u^0\partial_0\rho}{u(p+\rho)}+\frac{\partial_0u^0}{u}.
\end{equation}

In order to integrate Eqs. (\ref{trece}) and (\ref{quince}) over the radial
coordinate one must carefully choose what are the integration limits. In
principle, we would like to consider an observer placed on the region $%
r_{bh}\leq r\leq r_{c}$. Since there are two horizons, one can integrate
Eqs. (\ref{trece}) and (\ref{quince}) between two different pairs of limits:
i) from the cosmological horizon up to the observer, or ii) from the black
hole horizon up to the observer. In the first case one obtains,
\begin{equation}  \label{diecisiete}
ur^2\left(p+\rho\right)\left(u^2+\frac{\Delta}{r^2}\right)^{1/2}\exp\left(%
\int^r_{r_{c}}\mathrm{d}r f(r,t)\right)=C_1(t)
\end{equation}
and
\begin{equation}  \label{dieciocho}
r^2u\exp\left(\int^{\rho}_{\rho_{r_{c}}}\frac{\mathrm{d}\rho}{p+\rho}%
\right)\exp\left(\int^r_{r_{c}}\mathrm{d}r g(r,t)\right)=B_1(t),
\end{equation}
with $C_1(t)=-r_{c}^2\left(p+\rho\right)|_{r=r_c}$ and $B_1(t)=-r_{c}^2$,
where we have considered $u\rightarrow-1$ when $r\rightarrow r_c$. On the
other hand, if the lower integration limit is placed at the black hole
horizon, then we have
\begin{equation}  \label{diecinueve}
ur^2\left(p+\rho\right)\left(u^2+\frac{\Delta}{r^2}\right)^{1/2}\exp\left(%
\int^r_{r_{bh}}\mathrm{d}r f(r,t)\right)=C_2(t),
\end{equation}
\begin{equation}  \label{veinte}
r^2u\exp\left(\int^{\rho}_{\rho_{r_{bh}}}\frac{\mathrm{d}\rho}{p+\rho}%
\right)\exp\left(\int^r_{r_{bh}}\mathrm{d}r g(r,t)\right)=B_2(t),
\end{equation}
where $C_2(t)=-r_{bh}^2\left(p+\rho\right)|_{r=r_{bh}}$ and $B_2(t)=-r_{bh}^2
$, with a similar consideration on $u$ and $r$. Obviously, Eqs. (\ref%
{diecisiete}) and (\ref{diecinueve}) must be equivalent, as it should also
happen with Eqs. (\ref{dieciocho}) and (\ref{veinte}). In what follows it
will be seen that this is actually the case, because the same result is
obtained by evaluating Eq. (\ref{diecisiete}) (or (20)) at the black hole
horizon as Eq.(\ref{diecinueve}) (or (22)) at the cosmological horizon. It
will be also shown that the integral of the functions $f(r,t)$ and $g(r,t)$
must be finite as it should be expected.

Taking into account Eqs.(\ref{diecisiete}) and (\ref{dieciocho}) it can be
obtained,
\begin{widetext}
\begin{equation}\label{veintiuno}
\left(p+\rho\right)\left(u^2+\frac{\Delta}{r^2}\right)^{1/2}\exp\left(-\int^{\rho}_{\rho_{r_{c}}}\frac{{\rm
d}\rho}{p+\rho}\right)\exp\left(\int^r_{r_{c}}{\rm d}r
[f(r,t)-g(r,t)]\right)=\left(p+\rho\right)|_{r=r_c},
\end{equation}
\end{widetext}
and from Eqs. (\ref{diecinueve}) and (\ref{veinte}),
\begin{widetext}
\begin{equation}\label{veintidos}
\left(p+\rho\right)\left(u^2+\frac{\Delta}{r^2}\right)^{1/2}\exp\left(-\int^{\rho}_{\rho_{r_{bh}}}\frac{{\rm
d}\rho}{p+\rho}\right)\exp\left(\int^r_{r_{bh}}{\rm d}r
[f(r,t)-g(r,t)]\right)=\left(p+\rho\right)|_{r=r_{bh}}.
\end{equation}
\end{widetext}
Due to the spherical symmetry, the rate of black hole mass can be expressed
as $\dot{M}=4\pi r^2T^r_0$, or using Eqs. (\ref{dieciocho}) and (\ref%
{veintiuno}) or Eqs. (\ref{veinte}) and (\ref{veintidos}), as
\begin{equation}  \label{veintitres}
\dot{M}=4\pi r_{c}^2\left(p+\rho\right)|_{r=r_{c}}\exp\left(-\int^r_{r_{c}}%
\mathrm{d}r f(r,t)\right),
\end{equation}
or
\begin{equation}  \label{veinticuatro}
\dot{M}=4\pi
r_{bh}^2\left(p+\rho\right)|_{r=r_{bh}}\exp\left(-\int^r_{r_{bh}}\mathrm{d}r
f(r,t)\right).
\end{equation}
Thus, evaluating Eq.(\ref{veintitres}) at $r_{c}$ and Eq. (\ref{veinticuatro}%
) at $r_{bh}$, one easily obtains the black hole mass rate for observers at
the cosmological horizon and at the black hole horizon, respectively, which
turn out to be
\begin{equation}  \label{veinticinco}
\dot{M}=4\pi r_{c}^2\left(p+\rho\right)|_{r=r_{c}}
\end{equation}
and,
\begin{equation}  \label{veintiseis}
\dot{M}=4\pi r_{bh}^2\left(p+\rho\right)|_{r=r_{bh}}.
\end{equation}
Therefore, in order to obtain the black hole mass function in an isotropic
and homogeneous model one must integrate the following equation
\begin{equation}  \label{veintisiete}
\frac{\mathrm{d}M}{r_{c}^2}=4\pi\left(p+\rho\right)\mathrm{d}t ,
\end{equation}
for an observer at the cosmological horizon, and
\begin{equation}  \label{veintiocho}
\frac{\mathrm{d}M}{r_{bh}^2}=4\pi\left(p+\rho\right)\mathrm{d}t
\end{equation}
for an observer at the black hole horizon, which has the same
right-hand-side as in Eq.(\ref{veintisiete}). This can be easily integrated
in a quintessence model, where
\begin{equation}
\rho=\rho_0\left[a(t)/a_0\right]^{-3(1+w)}=\rho_0\left[1+\frac{3}{2}C
(1+w)(t-t_0)\right]^{-2} ,
\end{equation}
With $C$ now defined by $C=(8\pi\rho_0/3)^{1/2}$.

Using Eqs. (12) and (13), we can integrate the left-hand-side of Eqs. (29)
and (30) to give
\begin{equation}  \label{veintinueve}
-\frac{1}{2}\sqrt{\frac{\Lambda}{2}}\left(\frac{3+2 z}{\sqrt{1+z}} -\frac{%
3+2 z_0}{\sqrt{1+z_0}}\right)= \frac{4\pi\rho_0(1+w)(t-t_0)}{1+\frac{3}{2}%
(1+w)C(t-t_0)},
\end{equation}
where $z$ is given by,
\begin{equation}  \label{veintinueve_a}
z=\cos\left\{2/3\left[\arccos(-3M\Lambda^{1/2})+4\pi\right]\right\} ,
\end{equation}
in the first case, or by
\begin{equation}  \label{veintinueve_b}
z=\cos\left[2/3\arccos(-3M\Lambda^{1/2})\right] ,
\end{equation}
in the second one. Expressing the mass of the black hole in terms of $z$ by
the inverse of the previous expressions, i.e. using
\begin{widetext}
\begin{equation}\label{veintinueve_c}
x(z) \equiv 3\sqrt{\Lambda} \, M(z) = - \cos\left( \frac{3}{2}
\arccos z - 4 \pi \right)= - \cos\left( \frac{3}{2} \arccos z
\right) = (1 - 2 z) \sqrt{\frac{1 + z}{2}} ,
\end{equation}
\end{widetext}
one can note that the two different definitions of $z$, given by Eqs. (\ref%
{veintinueve_a}) and (\ref{veintinueve_b}), which imply two different
equations expressed through Eq.(\ref{veintinueve}), actually produce the
same mass function, i. e., the black hole mass function for an observer at
the cosmological horizon is the same that for an observer at the black hole
horizon. Therefore, one could assume that this is a well-defined problem and
that there is a unique black hole mass function independent of the placement
of the observer. On the other hand, because $z$ arises from an arccosine
function in Eq. (\ref{veintinueve_c}), it might at first sight range from $%
z=-1$ to $z=1$, but actually it has to take only on values in the interval $%
-1 < z < \frac{1}{2}$ to keep on non-negative values for the mass function (%
\ref{veintinueve_c}). In fact, $x(z)$ is an increasing function from $z=-1$
to $z=-\frac{1}{2}$, where it reaches the value $x=1$. It is worth noticing
that it is precisely the latter value the one at which both horizons
coincide. Thereafter, $x$ keeps being a decreasing function all the way
until $z=\frac{1}{2}$, where $x$ vanishes, i.e, when the black hole has
evaporated away.

As we have already mentioned above, the condition that allows us to have a
physical space between the two horizons is $3M\Lambda^{1/2}<1$. Therefore,
this condition may be considered to be also fulfilled at the current time
where the black hole has a mass $M_0$, i.e., when $x_0<1$, where $%
x=3M\Lambda^{1/2}$. In this way,at least in principle, a Taylor's expansion
could be a good treatment for Eq. (\ref{veintinueve}). In that case, one has
\begin{equation}  \label{treinta}
x\simeq\frac{x_0}{1-\frac{4}{3}x_0\sqrt{\frac{16\pi\rho_0}{3\Lambda}}F(t)},
\end{equation}
with
\begin{equation}  \label{treintayuno}
F(t)=\frac{3/2(1+w)C(t-t_0)}{1+3/2(1+w)C(t-t_0)}.
\end{equation}
Note that $F(t)$ would be an ever increasing function with time if $1+w>0$,
and a decreasing function with time if $1+w<0$. It follows then from Eqs. (%
\ref{veinticinco}) and (\ref{veintiseis}) that black holes would increase
their size if they accrete dark energy with $w>-1$, and would shrink if they
accrete phantom energy with $w<-1$. Therefore, in the phantom case $x$ will
take always on smaller and smaller values as the time grows up, making Eq.(%
\ref{treinta}) more and more accurate as the the black hole is evaporating
away to disappear at the big rip time
\begin{equation}  \label{treintaydos}
t_{br}=t_0+\frac{1}{(|w|-1)\sqrt{6\pi\rho_0}} .
\end{equation}

\begin{figure}[h]
\includegraphics[width=8cm]{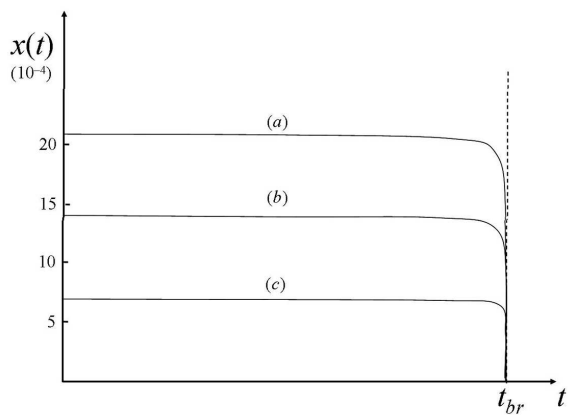}
\caption{Time evolution for the mass of a black hole within a
Schwarzschild-de Sitter universe filled with a phantom fluid ($w=-1.1$), as
evaluated by $x(t)=3\Lambda^{1/2}M(t)$, for different initial masses. From
top to bottom: $(a)$ $x_0=2.1 \, 10^{-3}$, $(b)$ $x_0 = 1.4 \, 10^{-3}$ and $%
(c)$ $x_0 = 7 \, 10^{-4}$, in units $H_0^{-1}$, i.e., $H_0=\frac{8 \protect%
\pi \protect\rho_0}{3} = 1$ and $H_0\lesssim \Lambda = 1.1$.}
\label{figura1}
\end{figure}

The previous line of reasoning is based on an approximate method. One could
also try to perform an analytical study in closed form by re-expressing Eq.(%
\ref{veintinueve}) as,
\begin{equation}  \label{treintaycuatro}
3 + 2 z = A(t) \sqrt{1+z} ,
\end{equation}
where the function $A(t)$ is given by,
\begin{equation}  \label{treintaycinco}
A(t) \equiv A_0 - \frac{c_0 (t - t_0) }{1+c_1 (t - t_0) } ,
\end{equation}
with
\begin{eqnarray}  \label{cuarentaytres}
A_0 &=& \frac{3+2 z_0}{\sqrt{1+z_0}} , \\
c_0 &=& 8 \pi \rho_0 \sqrt{\frac{2}{\Lambda}} \, (1+w) , \\
c_1 &=& \sqrt{6 \pi \rho_0} \, (1+w) ,
\end{eqnarray}
and $z_0 = \cos \left( \frac{2}{3} \left( \arccos(-3 M_0 \Lambda) + 4 \pi
\right) \right) $, $M_0$ being the initial mass of the black hole. If we
take $x_0\ll 1$, which is a sufficiently good approximation for any
practical initial mass of the black hole, then $z_0 \approx \frac{1}{2}$
(the value $z_0 = -1$, which corresponds to an initial black hole mass very
close to cero, is neglected in order to preserve a well-defined meaning for $%
A_0$).
\begin{figure}[h]
\includegraphics[width=8cm]{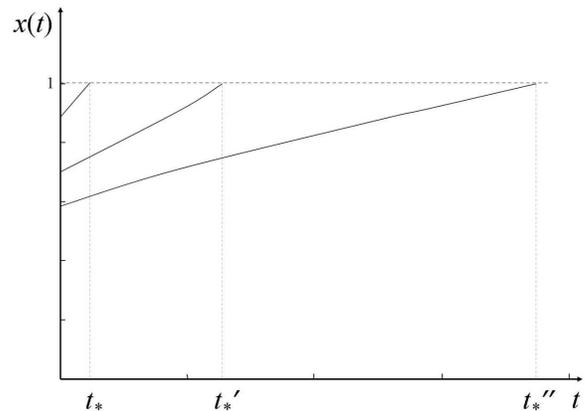}
\caption{Time evolution for the mass of the black hole within a
Schwarzschild-de Sitter universe filled with a dark energy fluid ($w=-0.9$)
as evaluated by $x(t)=3\Lambda^{1/2}M(t)$, for different values of the
initial mass: $x_0 = 0.88$, $x_0 = 0.70$ and $x_0 = 0.58$.}
\label{figura2}
\end{figure}

Squaring Eq.(\ref{treintaycuatro})we can attain the following two solutions,
\begin{equation}  \label{treintaynueve}
z_{\pm} = \frac{A^2(t) - 12 \pm |A(t)| \sqrt{A^2(t) - 8} }{8} .
\end{equation}
These two values, $z_\pm$, inserted in Eq. (\ref{veintinueve_c}), imply two
solutions for the time evolution of the black hole mass. The solution $z_+$
yields a negative value for the initial mass of the black hole, both in the
phantom and in the dark energy regime, so it ought to be disregarded. The
mass evolution for the value $z_-$ is depicted in Figs. 1-3. In the phantom
case, the time left until the big rip singularity turns out to be, $t_{br} -
t_0 = \frac{1}{|1+w|\sqrt{6 \pi \rho_0}}$. Then, from Eqs. (\ref%
{treintaycinco}) and (\ref{cuarentaytres}), it can be checked that as the
universe approaches the big rip, the function $A(t)$ tends to infinite,
i.e., $A(t\rightarrow t_{br}) \rightarrow \infty$, irrespective of the
initial mass of the black hole. Therefore, because in that limit, $z_-
\rightarrow -1$ and $x(z_-)\rightarrow 0$, any black hole eventually ends up
completely evaporated at the time in which the big rip singularity takes
place. That is the same conclusion as in Refs. \cite{Babichev:2004yx} and
\cite{MartinMoruno:2007se}. It follows that in the phantom case it does not
make any difference between considering the black hole in an asymptotically
flat scenario or in a cosmological scenario (see Fig. 1 which corresponds to
different values of the initial mass of the black hole). On the other hand,
Fig. 2, gives the case for dark energy, $w>-1$, where the mass of the black
hole grows up, and therefore the size of its event horizon also increases
whereas the cosmological horizon decreases until it eventually reaches the
cosmological one, at $x=1$.

\begin{figure}[h]
\includegraphics[width=8cm]{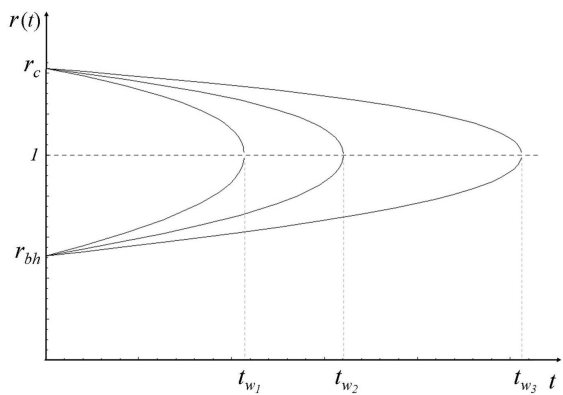}
\caption{Time evolution for both the black hole horizon, $r_{bh}$, and the
cosmological horizon, $r_c$, Eqs. (\protect\ref{diez}) and (\protect\ref%
{once}), for different values of the parameter $w$. From left to right: $w_1
= -0.88$, $w_2=-0.92$ and $w_3=-0.95$, with $A_0 =3.5$ and the same units as
those used in Fig. 1.}
\label{figura3}
\end{figure}
The time at which both the cosmological and the black hole horizons coincide
depends on the value of the parameter $w$,
\begin{equation}
t_w = t_0 + \frac{C}{1+w} ,
\end{equation}
where, $C=\frac{A_0 - 2\sqrt{2}}{8 \pi \rho_0 \frac{2}{\Lambda}+\sqrt{6 \pi
\rho_0}(A_0-2\sqrt{2})}$, is a constant. This case is depicted in Fig. 3. As
the value of $w$ approaches $-1$, the time at which the universe is engulfed
by the black hole grows up. Therefore, if we consider a cosmological
scenario, where the black hole space is asymptotically de Sitter a big hole
phenomenon would take place, which is so derived not only for the black hole
horizon growth but also because the cosmological horizon decreases. As it
should be expected in the limit of $w=-1$, the size of the black hole
remains constant in time, and the big hole phenomenon does not take place.

\section{Conclusions.}

In this work we have analyzed the accretion phenomenon of dark energy onto
non-static black holes in a consistent cosmic scenario. As a first approach,
we have considered a non-static Schwarzschild black hole onto which dark
energy is accreted in the presence of a positive cosmological constant. As
it was already shown in Ref. \cite{MartinMoruno:2007se}, upon accreating
phantom energy, any black hole decreases its mass and tends to disappear at
the big rip singularity. Coversely, if the parameter $w$ in the equation of
state is greater than $-1$, the black holes increase their sizes in a
moderate, safe way since they can never reach a size large enough to be able
to engulf the universe.

Then, in order to consider a more realistic cosmic scenario for
dark energy accretion, we have taken a non-static Schwarzschild-de
Sitter universe, in which the mass is an arbitrary function of
time. In this cosmological consistent framework, the accretion of
phantom energy onto black holes makes the size of the hole to
steadily decrease to finally vanish at the big rip time, such as
it is shown in Fig. 1. We would like to remark that the same
result was already obtained in Ref. \cite{Babichev:2004yx}
, where
this phenomenon was studied using a static Schwarzschild metric.
The reason for this agreement is that the accretion rate obtained
in the mentioned work may only be valid at small accretion rates,
a regime that is always satisfied in case of a phantom fluid.

For $w>-1$ in the mentioned cosmologically consistent scenario, accreating
black holes would increase their masses. The growth in the black hole mass
would produce an increase in the black hole horizon size, and a decrease in
the cosmological horizon size until the two horizon join up into just one,
when $M(t)=1/(3\Lambda^{1/2})$. Therefore, a big hole phenomenon would take
place in this case because the whole universe, delimited by the cosmological
horizon, is within the black hole horizon.

It is worth noticing that in this paper only a classical treatment has been
performed. It would be expected that quantum effects would probably smooth
these phenomena, leading to a non-vanishing black hole mass at the big rip
in the phantom case, \cite{Nojiri:2004pf}, and, in the case $w>-1$, to a
deceleration in the black hole horizon growth and in the cosmological
horizon decrease, due to particle creation at the two horizons induced by
the Unruh effect.

Finally, we would like to point out that in a recent work Gao et al. \cite%
{Gao:2008jv} declared that the mass of a black hole cannot decrease due to
phantom energy accretion, in contradiction with the results achieved in the
present work. We consider that their result is wrong because they introduced
a premise in which their previously desired result is actually contained,
making circular their whole argument. In particular, these authors choose a
particular function of the black hole mass proportional to the scale factor
at the beginning of section III and at the end of section IV. Obviously, if
one assumes a black hole mass function proportional to the scale factor, the
resulting black hole mass would only depend on the exponent of the scale
factor entering the mentioned function and on the time derivative of the
scale factor (which is positive for all expanding universes). On the other
hand, their initial assumption is totally contradictory with all studies on
dark energy accretion onto black holes in the literature for which the time
rate of the black hole function is recovered at least qualitatively at some
limit, such as it occurs in the present work.

\acknowledgements

This paper was supported by MEC under Research Project N%
${{}^o}$
FIS2005-01181, and by the Research Cooperation Project CSIC-CNRST. P.~M.-M.~
acknowledges CSIC and ESF for a I3P grant.

\end{document}